\documentclass[12pt]{article}
\usepackage{amsmath,amssymb}
\usepackage{graphicx,color,xcolor}
\usepackage{framed,multirow}
\usepackage{paralist,url}
\colorlet{framecolor}{black}
\colorlet{shadecolor}{lightgray}
\long\def\symbolfootnote[#1]#2{\begingroup%
\def\thefootnote{\fnsymbol{footnote}}\footnote[#1]{#2}\endgroup}
\def\brk#1{\left\langle#1\right\rangle}
\allowdisplaybreaks
\include{def}
\begin{document}
%%%%%%%%%%%%%%%%%%%%%%%%%%%%%%%%%%%%%%%%%%%

%%%%%%%%%%%%%%%%%%%%%%%%%%%%%%%%%%%%%%%%%%%
\thispagestyle{empty}
%%%%%%%%%%%%%%%%%%%%%%%%%%%%%%%%%%%%%%%%%%%
%\begin{flushright}
%\hfill{AEI-2014-xxx}
%\end{flushright}
%%%%%%%%%%%%%%%%%%%%%%%%%%%%%%%%%%%%%%%%%%%
\begin{center}

%\vspace{20pt}

{\Large\bf Fundamentals of the TianQin mission}

\vspace{10pt}

Jianwei Mei\symbolfootnote[1]{Email:~\sf jwmei@hust.edu.cn}, Chenggang Shao and Yan Wang

\vspace{10pt}

{\it\small
\baselineskip 20pt
{MoE Key Laboratory of Fundamental Quantities Measurement, School of Physics,\\[-5pt]
Huazhong University of Science and Technology, 1037 Luoyu Rd., Wuhan 430074, P.R. China}
}

\vspace{1cm}
{\bf Abstract}
\end{center}
\vspace{-10pt}

This talk adds more detail to the proposed TianQin mission.\cite{TQ} The response of TianQin to a binary source is an important problem for the experiment. Here we present the detail of the response function of an equal-arm Michelson interferometer to the gravitation waves of a binary source with a circular orbit.

%%%%%%%%%%%%%%%%%%%%%%%%%%%%%%%
% \newpage
% \setcounter{footnote}{0}
% \setcounter{page}{1}
%%%%%%%%%%%%%%%%%%%%%%%%%%%%%%%

%\tableofcontents
%%%%%%%%%%%%%%%%%%%%%%%%%%%%%%%

\section{Introduction}

General transfer (response) functions for LISA-like missions has been studied in much detail in the past (see, e.g. \cite{Cornish:2002rt}). Here we adapt the results to the particular case of TianQin and present the detailed form of the transfer function used in our calculations \cite{TQ}, assuming that the source is a binary system with a circular orbit.

\section{Gravitational waves from a binary source}

\begin{figure}
\begin{center}
\includegraphics[width=8cm,height=6cm]{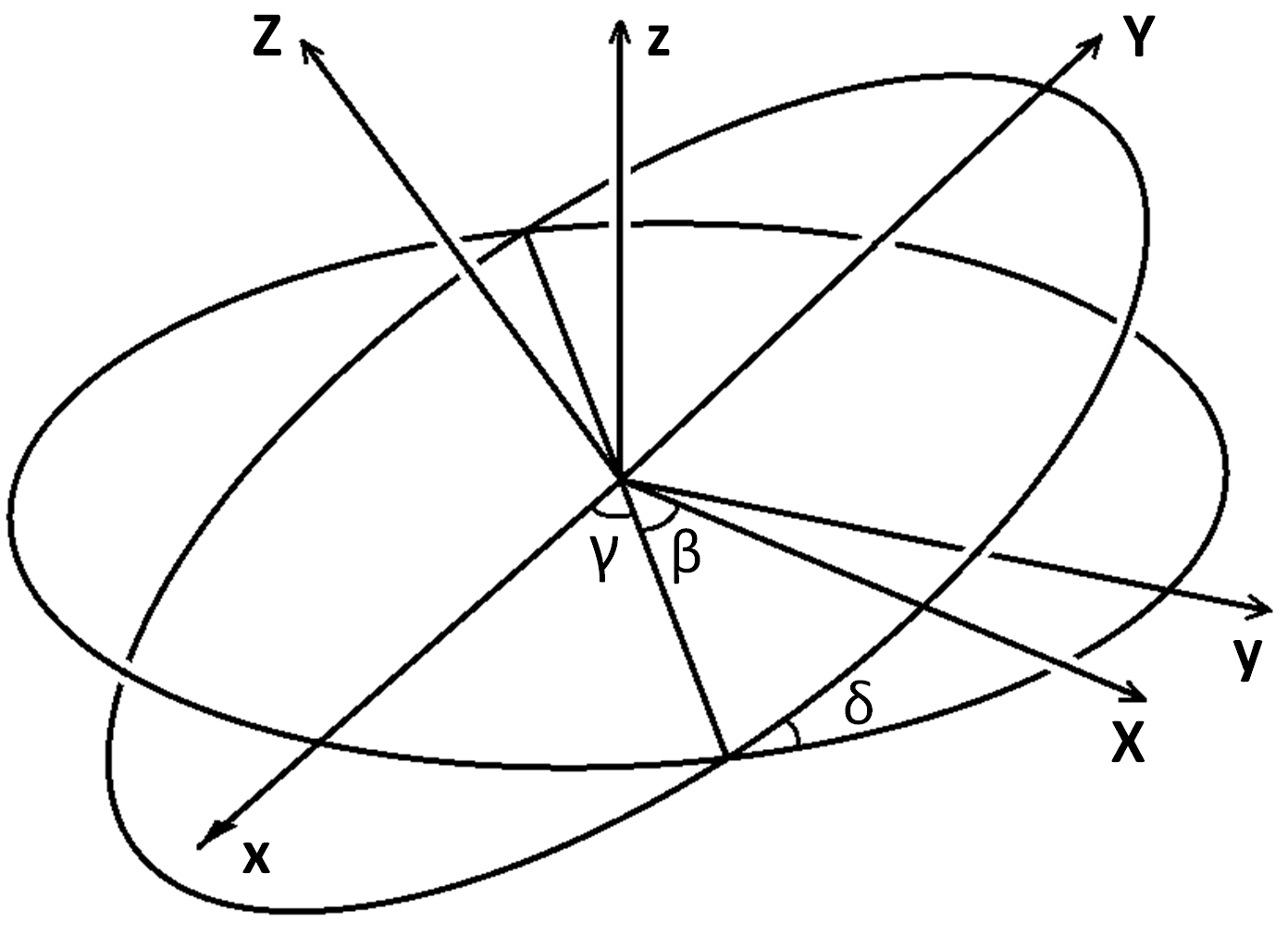}
\caption{The coordinates $(X,Y,Z)$ are adapted to the binary source: $\hat{\bf X}$ points along the semi-major axis, $\hat{\bf Y}$ points along the semi-minor axis and $\hat{\bf Z}$ is perpendicular to the orbital plane. The coordinates $(x,y,z)$ also originate from the barycenter of the source, with $\hat{\bf z}$ pointing to the detector; $\hat{\bf x}$ and $\hat{\bf y}$ can be chosen as needed.}
\label{fig.binary}
\end{center}
\end{figure}

To specify the orientation of the binary orbit and the relative direction between the detector and the source, we follow \cite{Poisson2014} and set up the coordinate systems as in Fig. \ref{fig.binary}.

For a binary source with a circular orbit, the gravitational waves can be written as
%%%
\begin{eqnarray} h_+&=&h_0a_+e^{2iwt} +c.c.\,,\quad a_+=-\Big[ \frac12(1+c_\delta^2) c_{2\gamma}+ic_\delta s_{2\gamma}\Big] e^{2i\beta}\,,\nonumber\\
h_\times&=&h_0a_\times e^{2iwt}+c.c.\,,\quad a_\times=-\Big[\frac12 (1+c_\delta^2) s_{2\gamma} -ic_\delta c_{2\gamma} \Big] e^{2i\beta}\,,\label{h.J0806}\end{eqnarray}
%%%
where $h_+$ and $h_\times$ are the two polarizations of the gravitational waves; $w$ is the angular orbital frequency of the source;  $h_0=\frac{2 G_N^2M_1M_2}{Da}\,$, with $M_1$ and $M_2$ being the masses of the two component stars, $D$ being their distance to the Earth, and $a=[G_N(M_1+M_2) /w^2]^{1/3}$ is the length of the semi-major axis; $c_\delta\equiv \cos\delta\,$, $s_\delta \equiv\sin\delta\,$, and so on; the angles $\beta\,$, $\gamma$ and $\delta$ are indicated in Fig. \ref{fig.binary}. Note we take the speed of light $c=1\,$.

TianQin is currently using the binary system J0806.3+1527 as a tentative source for gravitational waves. In this case, we find $h_0\approx6.4 \times10^{-23} \Big(\frac{ M_1}{0.55 M_\odot} \Big)\Big(\frac{M_2}{0.27 M_\odot}\Big)\Big(\frac{ 5\,{\rm kpc}}{D}\Big) \Big(\frac{6.6\times 10^4\,{\rm km}}{a}\Big)\,$.

\section{The transfer function}

Our calculation of the transfer function follows that of \cite{Larson:1999we}. To save space, we will only present the key steps here. The most important difference is that we will include the contribution from both polarizations of the gravitational waves from the binary source. In this case, the total power on the laser interferometer is, instead of (21) in \cite{Larson:1999we},
%%%
\begin{eqnarray}\brk{\Delta^2}&=&\int_0^{+\infty}\frac{dw}{2\pi}\,\frac{h_1^2T_1(w) +h_2^2T_2(w) -2h_1h_2T_3(w)}{(w\tau)^2}\,,\nonumber\\
h_i&=&h_+(w)\cos2\tilde\phi_i+h_\times(w)\sin2\tilde\phi_i\,,\nonumber\\
T_1&=&\mu_1^2\Big[1+\cos^2(w\tau)-2\cos(w\tau)\cos(\mu_1w\tau)\Big]\nonumber\\
&&-2\mu_1\sin(w\tau)\sin(\mu_1w\tau)+\sin^2(w\tau)\,,\nonumber\\
T_2&=&\mu_2^2\Big[1+\cos^2(w\tau)-2\cos(w\tau)\cos(\mu_2w\tau)\Big]\nonumber\\
&&-2\mu_2\sin(w\tau)\sin(\mu_2w\tau)+\sin^2(w\tau)\,,\nonumber\\
T_3&=&\mu_1\mu_2\Big[\cos(w\tau)-\cos(\mu_1w\tau)\Big]\Big[\cos(w\tau) -\cos(\mu_2w\tau)\Big]\nonumber\\
&&+\Big[\sin(w\tau)-\mu_1\sin(\mu_1w\tau)\Big]\Big[\sin(w\tau) -\mu_2\sin(\mu_2w\tau)\Big]\,,
\label{def.delta2}\end{eqnarray}
%%%
where $\mu_i=\cos\theta_i\,$. Note that the $i$'th laser link is at an angle $\theta_i$ with respect to $\hat{z}\,$, and at an angle $\tilde\phi_i$ with respect to $\hat{x}\,$; $c\,\tau$ is the arm length of the laser interferometer. More detail of the definition of $\brk{\Delta^2}$ can be found in \cite{Larson:1999we}.

Plugging the detailed expression (\ref{h.J0806}) of the gravitational waves into (\ref{def.delta2}), one can write $\brk{\Delta^2} =\frac1{2\pi} \int_0^{+\infty}h_0^2R_0(w)dw\,$, in contrast to
(16) and (28) in [39] in \cite{Larson:1999we}, with
%%%
\begin{eqnarray} R_0(w)&=&\Big|a_+\cos2\tilde\phi_1+a_\times\sin2\tilde\phi_1\Big|^2 \frac{T_1 (w)}{(w\tau)^2}\nonumber\\
&&+\Big|a_+\cos2\tilde\phi_2+a_\times \sin2\tilde\phi_2\Big|^2 \frac{T_2(w)}{(w\tau)^2}\nonumber\\
&&-2\Re e\Big[\Big(a_+\cos2\tilde\phi_1 +a_\times\sin2\tilde\phi_1\Big) \nonumber\\
&&\qquad\times\Big(a_+\cos2\tilde\phi_2 +a_\times\sin2\tilde\phi_2\Big) \Big] \frac{T_3(w)}{(w\tau)^2}\,.\end{eqnarray}
%%%

For TianQin, $\theta_1=\theta_2=90^\circ\,$ (hence $\mu_1 =\mu_2 =0\,$) and $\tilde\phi_2=\tilde\phi_1+\pi/3\,$. We find in the low frequency limit ($w\tau\to0\,$),
%%%
\begin{eqnarray} R_0(w)&\approx&\Big|\frac12(1+c_\delta^2)\Big(c_{2(\gamma-\tilde\phi_1)} -c_{2(\gamma-\tilde\phi_1-\pi/3)}\Big)\nonumber\\
&&+ic_\delta\Big(s_{2(\gamma-\tilde\phi_1)} -s_{2(\gamma-\tilde\phi_1-\pi/3)}\Big)\Big|^2\,.\end{eqnarray}
%%%
The parameter $\gamma$ is difficult to measure for a distant astrophysical object, and $\tilde\phi_1$ varies over time due to the orbital motion of satellites. Averaging (numerically) over $\gamma$ and $\tilde\phi_1\,$, we obtain
%%%
\begin{equation} R_0(w)\approx3-2.6\sin^2\delta\,,\label{val.R0} \end{equation}
%%%
which is the formula used in \cite{TQ}.


\begin{thebibliography}{99}

\bibitem{TQ} J. Luo, {\it et al}, ``TianQin: a space-borne gravitational wave detector", Submitted to Class. Quant. Grav..

\bibitem{Cornish:2002rt}
  N.~J.~Cornish and L.~J.~Rubbo,
  %``The LISA response function,''
  Phys.\ Rev.\ D {\bf 67}, 022001 (2003)
  [Phys.\ Rev.\ D {\bf 67}, 029905 (2003)]
  [gr-qc/0209011].

\bibitem{Poisson2014} E. Poisson and C. M. Will,
  {\it Gravity: Newtonian, Post-Newtonian, Relativistic},
  Cambridge University Press (29 May 2014)

\bibitem{Larson:1999we}
  S.~L.~Larson, W.~A.~Hiscock and R.~W.~Hellings,
  %``Sensitivity curves for spaceborne gravitational wave interferometers,''
  Phys.\ Rev.\ D {\bf 62}, 062001 (2000)
  [gr-qc/9909080].


\end{thebibliography}
\end{document}